# A Supersolid Skin Covering both Water and Ice

Xi Zhang[1,2,a], Yongli Huang[3,a], Zengsheng Ma[3], Yichun Zhou[3,*], Weitao Zheng[4], Ji Zhou[5], and Chang Q Sun[1-5,*]

[1] *NOVITAS, School of Electrical and Electronic Engineering, Nanyang Technological University, Singapore 639798*

[2] *Center for Coordination Bond and Electronic Engineering, College of Materials Science and Engineering, China Jiliang University, Hangzhou 310018, China*

[3] *Key Laboratory of Low-dimensional Materials and Application Technology (Ministry of Education) and Faculty of Materials, Optoelectronics and Physics, Xiangtan University, Xiangtan, 411105, China*

[4] *School of Materials Science, Jilin University, Changchun 130012, China*

[5] *State Key Laboratory of New Ceramics and Fine Processing, Department of Materials Science and Engineering, Tsinghua University, Beijing 100084, China*

[a] X.Z. and Y.H. contribute equally.

## Contents

Abstract

The mysterous nature and functionality of wter and ice skins remain baffling to the comnminity since 1859 when Farady firstly proposed liquid skin lubricating ice. Here we show the presence of supersolid phase that covers both water and ice using Raman spectrscopy measurments and quantum calculations. In the skin of two molecular layers thick, molecular undercoordination shortens the H-O bond by ~16% and lengthens the O:H nonbond by ~25% through repulsion between electron pairs on adjascent O atoms, which depresses the density from 0.92 for bulk ice to 0.75 g·cm$^{-3}$. The O:H-O cooperative relaxation stiffens the H-O stretching phonon from 3200/3150 cm$^{-1}$ to the same value of 3450 cm$^{-1}$ and raises the melting temnperature of both skins by up to ~310 K. Numerical derivatives on the viscoecity and charge accuumulation suggests that the elastic, polarized, and thermally stable supersolid phase makes the ice frictionless and water skin hydrophobic and ice like at room temperature.

TOC entry (see video clip as supportinginformation)

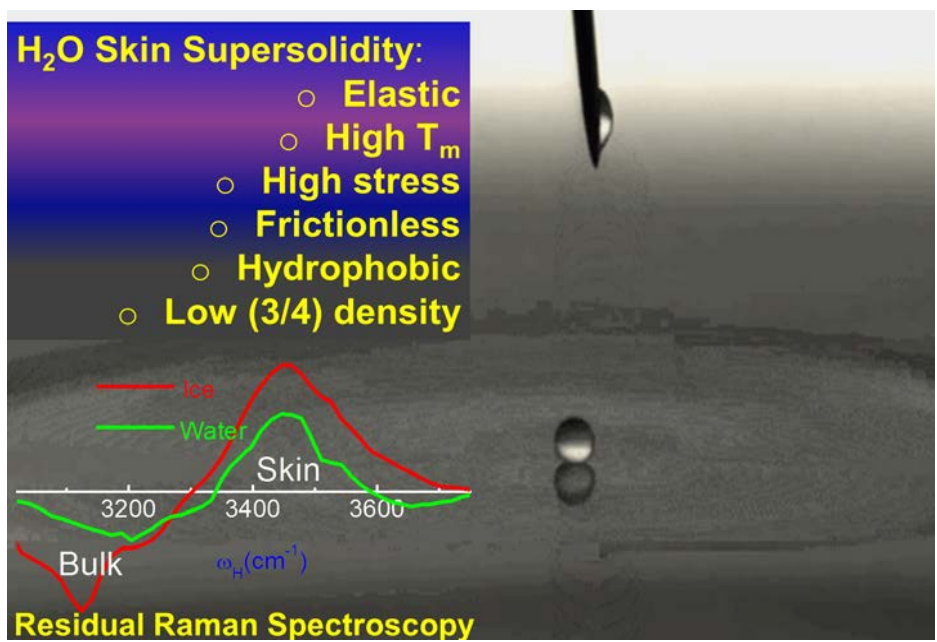

1. **Introduction**

Ice skin is most slipper of known[1-5] and the surface tension of water, 72 dynes/cm at 25 °C[6], is extraordinarily high. Ice remains slippery while one is standing still on it without even pressure melting or friction heating. Small insects such as a water strider can walk and glide freely on water because: (i) it weighs insufficiently to penetrate the skin and (ii) the interface between its paddle and the skin of water is hydrophobic. If carefully placed on the surface, a small needle floats on the water even though its density is times higher than that of water. If the surface is agitated to break up the tension, then the needle will sink quickly. A high-speed video clip (see Supplementary Information) shows that the repeatedly bouncing of a water droplet on water surface, evidencing that water skin is elastic and hydrophobic.

The high tension and hydrophobicity of water surface was commonly attributed to the presence of a layer of solid ice[7-9] while ice slippery was perceived as the presence of a liquid or a quasi-liquid skin serving as the lubricant[10] even at temperature below freezing[11,12]. Mechanisms of pressure-promoted melting[13] and friction-induced heating[14] also explained the slippery of ice.

However, the pressure-promoted melting contradicts with the phase diagram of ice below -22 °C, i.e., pressure induces solid-solid transition rather than solid-liquid transition[15], though the pressure could lower the melting point of ice[13]. It is expected that at melting the vibration amplitudes of the surface molecules are folds greater than that in the bulk but an interfacial force microscopy measurements ruled out this prediction[2]. The skin layer is viscoelastic at temperatures ranging from -10 to -30 °C, which evidences the absence of the liquid layer at such low temperatures. The surface pre-melting is also in conflicting with the ice-like nature of ultrathin films of water[8,9,16]. Vibrational sum frequency spectroscopy measurements and molecular dynamics (MD) calculations suggested that the outermost two layers of water molecules are ordered "ice like" at room temperature[17,18], being opposite to the premelting of surface or defect sites demonstrated by other usual materials such as metals[19].

Amazingly, water freezing starts preferentially at sites of molecules with full-coordination neighbors of four. According to MD calculations[20], freezing starts firstly in the subsurface of water instead of the top layer that remains disordered during freezing. The bulk melting is mediated by topological defects that preserve the coordination environment of the tetrahedral-coordinated network. Such defects form a region with a longer lifetime that the ideal bulk[21]. Water droplet on roughened Ag surface (with nanocolumnar structures) having a greater contact angle melts 62 sec later than the latter at -4 °C compared with the droplet with smaller contact angle on a smooth Ag surface[22]. The critical temperature for transiting the contact the initial contact angle of a droplet to zero is proportional to the its initial value (curvature)[23]. Transition happens at 185, 234, and 271 °C for water droplets on quartz, sapphire, and graphite with initial contact angles of 27.9, 64.2, and 84.7°, respectively. These findings indicate that water molecules at highly curved skin are thermally even more stable.

Why is ice so slippery and why is the tension of water skin so high? Does water form skin on ice or the alternative? These questions have baffled the community since 1859 when Farady[12] firstly proposed the presence of a lubricant liquid skin on ice.

We aim to show in this communication that molecular undercoordination induced H-O bond contraction and the O:H nonbond elongation and the associated core electron entrapment and non-bonding lone-pair polarization[24-26] result in the high-elasticity[4], self-lubricavity of ice[1,27] and the hydrophobicity[16,28] of water and ice skin. Ice slippery results from the lone pair weak yet elastic interaction, which makes the ice self lubricate. The cohesive energy gain of the shortened O-H bond raises the critical temperature of phase transition[26], and therefore, a monolayer of water performs like ice at room temperature with high elasticity[29] and high core charge density[30-32]. The monolayer skin melts at temperature about 310 K according to the MD derivatives[33].

## 2 Principle: O:H-O bond cooperativity

Firstly, hydrogen bond (O:H-O) relaxes cooperatively under excitation. The O:H-O bond forms a pair of asymmetric, coupled, H-bridged oscillators with ultra-short interactions[34,35]. The cooperativity of the O:H-O bond discriminates water ice from other usual materials in the structure order and physical properties. In practice, O atoms always move in the same direction along the O:H-O bond by different amounts[26]. The softer O:H ($d_L$ length) bond always relaxes more in length than the stiffer H-O ($d_H$) covalent bond does: $\left|\Delta d_L\right| > \left|\Delta d_H\right|$. The relaxation in length and energy and the associated local charge distribution of the O:H-O bond determine the anomalies of water ice under various stimuli such as compressing[26,36,37], clustering[24], and cooling[25,38]. With the given structure order of tetrahedrally-coordinated molecules and known mass density ρ, one is able to determine the molecular size ($d_H$, unit in Å), separation ($d_{OO}$ or $d_L$), of molecules packing in water and ice, in terms of statistic mean[34],

$$\begin{cases} d_{oo} = 2.6950\rho^{-1/3} & (Molecular\ separation) \\ \dfrac{d_L}{d_{L0}} = \dfrac{2}{1+exp\left[(d_H - d_{H0})/0.2428\right]}; & (d_{H0}=1.0004\ and\ d_{L0}=1.6946\ at\ 4\ °C) \end{cases} \quad 1$$

Secondly, the impact of the O:H-O bond cooperative relaxation is tremendous. The H-O bond contracts and the the O:H nonbond expands spontenously once the molecular coordination number (CN) is reduced from the ideal bulk value of four, which happens in skins defect sites, hydrotion shells, and ultrathin films [24]. The shortended H-O bond becomes stiffer with a depression of the interatomic potential well, which entraps and densifies the O 1s binding energy[31,39-41] and local core and bondinh charge [27,32,39,42,43]. The stiffening of the H-O bond raises the critical tempertaure for phase transition and stiffens the H-O stretching $\omega_H$ phonons. Most importantly, the densely entrapped bonding charge polarizes the soluted nonbonding electrons with bound energy (equivalent of work function) changing from 3.2 in the bulk to 1.6 eV in the skin [27]. The large O:H bond elongation partly compensates for the H-O contracion and partly lengthens the overall O—O distance, resultinfg in volume expansion[44-47]. The measured 5.9%~10% skin $d_{OO}$ expansion at the ambinet [34,46] does follow the expectation. Likewise, the 4.0% and 7.5% volume expansion of water molecules confined, respectively, in the 5.1 and 2.8 nm sized hydrophobic $TiO_2$ pores[48] provide further evidence for the predicted low density skin.

### 3. Experimental and numerical verification

In order to verify our predictions, we analyzed the nice set of Raman data collected by Donaldson and co-workers[4] and conducted quantum calculations using the DFT-derived MD and the dispersion-corrected DFT packages. Calculations was focused on resolving the O:H-O cooperative relaxation in length and stiffness in the skin and the skin charge accumulation. MD calculations were performed using the Forcite code with COMPASS forcefield.[49] Ice interface is relaxed in NPT ensemble at 180 K for 100 ps to obtain equilibrium. The time step is 0.5 fs. Nose-hoover thermostat with Q ratio of 0.01 is adopted to control the temperature.

Dispersion-corrected DFT structural optimizations of ice surface were performed using $Dmol^3$ code based on the PBE functional[50] in the general gradient approximation and the dispersion-corrected Tkatchenko-Scheffler scheme[51] with the inclusion of hydrogen bonding and vdW interactions. The all-electron method was used to describe the wave functions with a double numeric and polarization basis sets. The self-consistency threshold of total energy was set at $10^{-6}$ Hartree. In the structural optimization, the tolerance limits for the energy, force and displacement were set at $10^{-5}$ Hartree, 0.002 Hartree/Å and 0.005 Å, respectively.

Calculations also extend to examine the charge densification, viscocity, and surface tension at the skin region of 200 K ice. The conventionally used super-cell[52] as shown in Figure 1 was used in calculations. A vacuum slab is inserted into the super-cell to approximate the liquid-vapor interface or the skin that contains free H-O radicals.

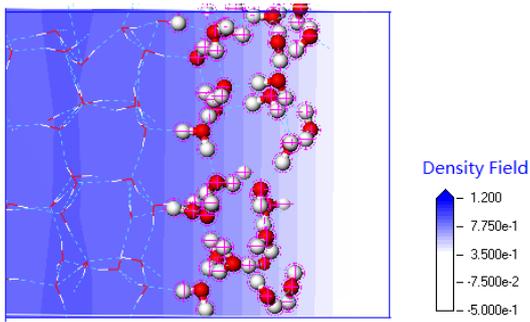

Figure 1 $H_2O$ super-cell with a vacuum slab representing the supersolid skin of ice (200 K) with even undercoordinated H-O radicals with color indicating the MD-derived density field. The current work turns out the skin reduces the $H_2O$ molecular volume but inflates molecular distance because molecular undercoordination shortens and stiffens the $d_H$ and lengthens and polarizes the $d_L$ due to repulsion between the electron pairs on adjacent O atoms and the O:H-O bond mechanical strength disparity.

## 4. Results and discussion

### *4.1 Skin-resolved O:H-O bond relaxation*

Indeed, the O:H-O bond cooperative relaxation lowers the skin mass density substantially. Figure 2 features the residual length spectra (RLS) for the MD-derived O-H and O:H bond of ice. The RLS is obtained by subtracting the spectrum of a fully-filled unit cell from that containing the vacuum slab. The RLS turns out that the O-H bond contracts from ~1.00 in the body to ~0.95 at the skin. The O:H elongates from ~1.68 in the body to ~1.90 Å (average) at the skin, resulting in a 6.8% $d_{OO}$ elongation or a 20% volume expansion. The $d_H$ = 0.93 Å peak and the broaden $d_L$ could be reflection of the even undercoordinated H-O radicals that are associated 3650 cm$^{-1}$ vibration frequency[24]. This result agrees with the trend that the H-O bond contracts from 0.9732 Å at the center to 0.9659 Å at the skin of a water droplet containing 1000 molecules in previous MD calculations[53].

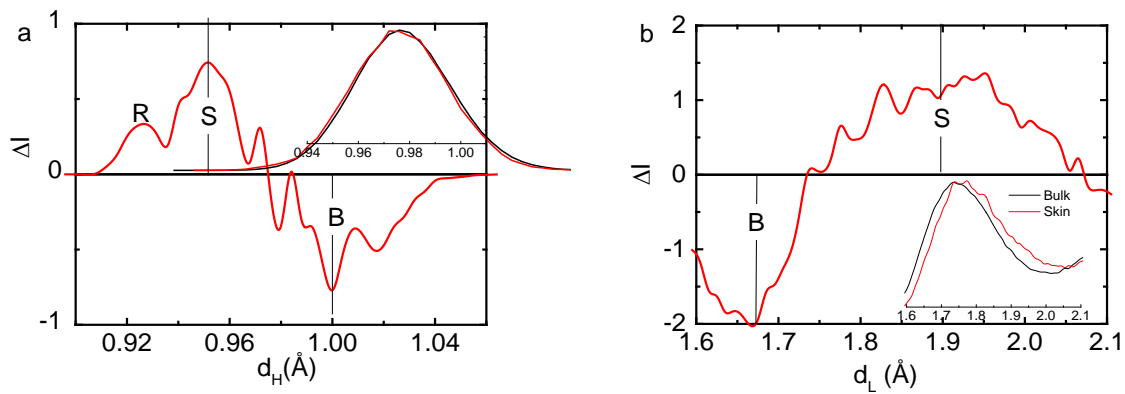

Figure 2 MD-derived RLS of (a) $d_H$ contraction from ~1.00 to ~0.95 (0.93 for radicals) Å and (b) $d_L$ elongation from ~1.68 to ~1.90 Å (high fluctuation). Insets show the raw data of calculations. R, S and B represent the H-O radicals, skin, and bulk.

Furthermore, the measured $d_{OO}$ (= 2.965 Å) [46] in the water skin turns out, using eq (1), $d_H$ = 0.8406 Å and $d_L$ = 2.1126 Å, corresponding to the 0.75 g·cm$^{-3}$ mass density that is much lower than the bulk ice of 0.92 g·cm$^{-3}$. Therefore, the numerical RLS and the experimental derivatives verified the expectation of the ultra-low density of the skins of both water and ice. Table 1 lists the experimentally-derived bond length ($d_{OO}$, $d_x$) and density (ρ) (eq 1), vibration frequency ($\omega_x$), and bond energy ($E_x$) that was transformed from Lagrangian solution to the O:H-O oscillators with the measured $d_{OO}$ and $\omega_x$ as input[35]. Clearly, skins of water and ice share the same identity of density, bond length, bond strength, vibration frequency, etc., at temperatures of 25 and -20 °C.

Table 1 Skin supersolidity ($\omega_x$, $d_x$, $E_x$, ρ) of water and ice converted from eq (1) and Lagrangian solution with the known data (indicated with refs) as input based on the water structure[34,35]. $q_H$ is the effective charge of the proton, which is an adjustable for testing.

|  | Water (298 K) | | Ice ($\rho_{min}$) | Ice | Vapor |
|---|---|---|---|---|---|
|  | bulk | skin | bulk | 80 K | dimer |
| $\omega_H$(cm$^{-1}$) | 3200[4] | 3450[4] | 3125[4] | 3090[25] | 3650[54] |
| $\omega_L$(cm$^{-1}$)[25] | 220 | ~180[24] | 210 | 235 | 0 |
| $d_{OO}$(Å) [34] | 2.700[55] | 2.965[46] | 2.771 | 2.751 | 2.980[46] |
| $d_H$(Å) [34] | 0.9981 | 0.8406 | 0.9676 | 0.9771 | 0.8030 |
| $d_L$(Å) [34] | 1.6969 | 2.1126 | 1.8034 | 1.7739 | ≥2.177 |
| ρ(g·cm$^{-3}$) [34] | 0.9945 | 0.7509 | 0.92[56] | 0.94[56] | ≤0.7396 |
| $E_L$(meV)∝$(\omega_x \times d_x)^2$ | 91.6 | 95[57] | 94.2 | 114.2 | 0 |
| $E_L$(meV) ($q_H$ = 0.20 e) | 24.6 | 24.4 | 26.2 | 44.3 | 0 |
| $E_L$(meV) (0.17 e) | 33.4 | 33.8 | 35.1 | 52.0 | 0 |
| *$E_H$(eV) ($q_H$ = 0.20 e) | 3.6201 | 7.1967 | 4.0987; 3.97[25] | 3.9416 | 8.6429 |
| $E_H$(eV) (0.10 e) | 3.6203 | 7.1968 | 4.0990 | 3.9418 | 8.6429 |

*4.2 Phonon frequency and thermal stability*

The supersolid skins of water and ice stiffen the $\omega_H$ to the identical frequency of 3450 cm$^{-1}$ and raise the melting temperature ($T_m$) substantially. The phonon frequency corresponds to the square root of the stiffness of the specific bond with $Y_x$ being the Young's modulus and the $T_m$ is proportional to the $E_H$[24,35]:

$$\begin{cases} \Delta\omega_x \propto \sqrt{(k_x + k_c)/\mu_x} \propto \sqrt{E_x/\mu_x}/d_x = \sqrt{Y_x d_x/\mu_x} \\ T_m \propto E_H \end{cases} \quad (T_m < T_{evap})$$

(2)

Where $k_x$ and $k_c$ are the force constants, being the 2nd differential of the respective potential (exchange for H-O, van der Waals for O:H, and Coulomb potential for the O—O). The $\mu_x$ is the reduced mass of respective $H_2O:H_2O$ ($\omega_L$) and the H-O ($\omega_H$) vibrating dimer.

As demonstrated in [24], the temperature of melting is proportional to the H-O bond energy. The stiffening of the H-O bond raises the local freezing/melting temperature accordingly. Therefore, the supersolid skin of water performs like ice at room temperature. This derivative also applys to water droplets encapsulated in hydrophobic nanopores[29,66,67] and defects elavated local $T_m$ [20,21]. According to the $E_H$ derived in Table 1, the skin $T_m$ is twice that of the bulk, but the $T_m$ is subject to the temperature of evaporation $T_{evap}$ that requires energy breaking the H:O bond (0.925 eV[57]). Therefore, the monolayer water melts at temperature about 315 K[33].

Figure 3(a, b) features the calculated residual phonon spectra (RPS) of $\omega_L$ and $\omega_H$ of ice in comparison to (c) the $\omega_H$ measurements from water and ice with insets showing the original raw data [4]. The RPS were obtained by subtracting the spectrum collected at smaller angles (between the surface normal and the reflection beam) from the one collected at larger angles upon the spectral area being normalized. The valleys of the RPS represent the bulk feature while the peaks the skin attributes. Proper offset of the calculated RPS is necessary as the calculation code overestimates the inter- and inter-molecular interactions[25].

As expected, the $\omega_L$ undergoes a redshift while the $\omega_H$ splits into three peaks at the skin. Two $\omega_H$ of higher frequencies arise from the undercoordination induced contraction of the bonded (T) and the free H-O radicals (R). The $\omega_L$ redshift arises from O-O repulsion and polarization. The polarization in turn screens and splits the interatomic potential, resulting in another $\omega_H$ peak (P) below the bulk valley. The R peak corresponds to the even lower-coordinated H-O radicals.

Most strikingly, the measured RPS shows that skins of water and ice share the identical $\omega_H$ at 3450 cm$^{-1}$, which indicates that both skins are the same in nature. This finding agrees with the DFT-MD derivative that the H-O vibration frequency shifts from ~3250 cm$^{-1}$ at a 7 Å depth to ~3500 cm$^{-1}$ at the 2 Å skin of water liquid [52]. Therefore, neither ice skin forms on water nor does the otherwise. One can refer such abnormal skins to the supersolid state. The concept of supersolidity is adopted from the superfluidity of solid $^4$He at mK temperatures. The skins of $^4$He fragments are highly elastic and frictionless with repulsive force between them at motion[58].

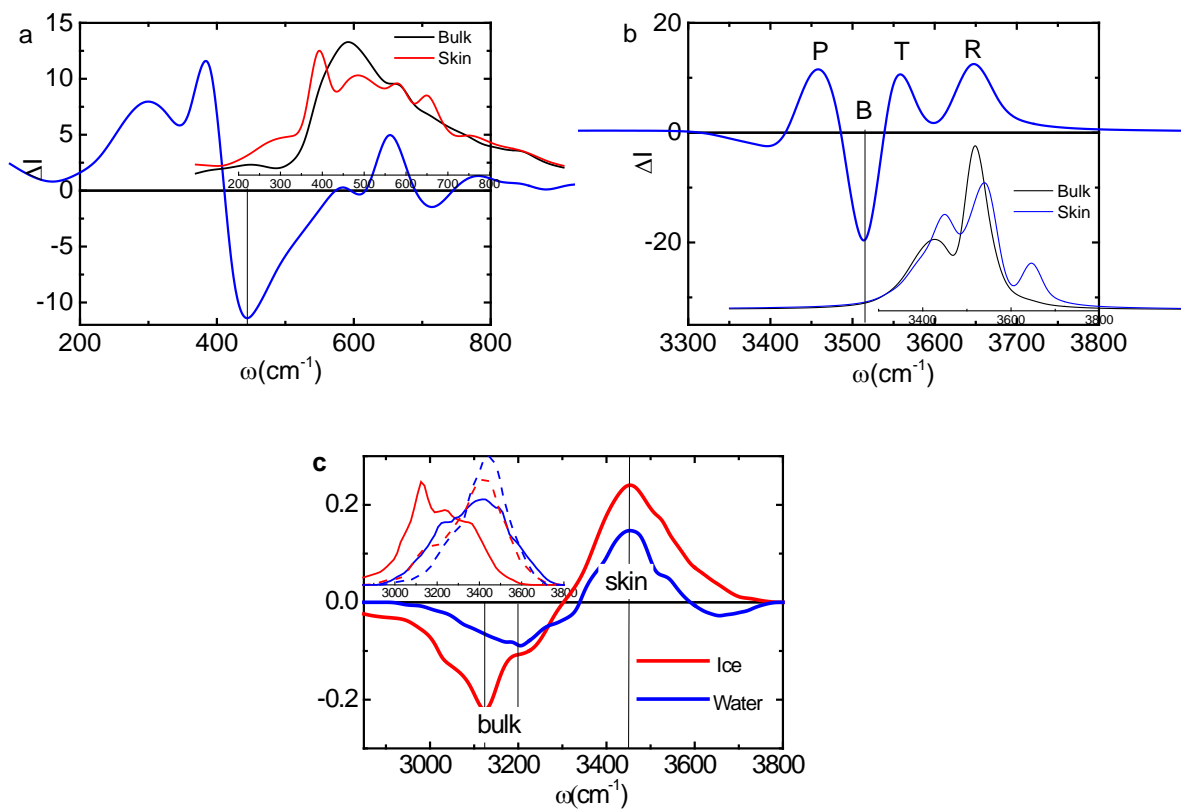

Figure 3 RPS of the MD-derived (a) $\omega_L$ and (b) $\omega_H$ of ice and (c) the measured $\omega_H$ of water (at room-temperature) and ice (at -20 and -15 °C) [4]. The $\omega_L$ undergoes redshift while the $\omega_H$ splits into three. Features T and R correspond to the H-O skin bond and the free H-O radicals and P to the screen and splitting effect on the crystal potentials by the undercoordinated O:H in numerical derivatives. Water and ice skins share the identical $\omega_H$ of 3450 cm$^{-1}$, which clarifies neither liquid skin forms on ice nor solid skin on water.

*4.3 Skin charge entrapment and polarization-repulsive forces*

The skins do entrap bonding and core electrons, which polarizes nonbonding charge, raises the thermal stability, and creates electronic repulsive force. Table 2 features the DFT-MD derived $d_x$ and the charge polarization at the skin of water. The net charge of a water molecule increases from 0.022 in the body to -0.024 at the skin. As expected, bonding electron gain does happen at the skin. The charge gain and the nonbonding electron polarization provide the sources for the Coulomb repulsive force that reduces the friction at the interface between ice and other materials. Measurements have revealed the presence of the repulsive forces between a hydrated mica-tungsten contacting pair at 24 °C[59]. Such repulsive interactions appear at above 20% relative humidity (RH) and are fully developed in the range of 38-45% RH. The repulsion corresponds to an elastic modulus of 6.7 GPa. Monolayer ice also forms on graphite surface at 25% RH and room temperature[60,61]. These experimental and numerical findings evidence the presence of the supersolidity with repulsive forces as a result of surface polarization and bonding charge accumulation and the elevated melting point resulting due to H-O strength gain.

Table 2 Numerically-derived O:H-O length relaxation, charge densification and polarization at the skin of ice. Negative sign represents net electron gain. MD and DFT show the same trend despite the quantitative discrepancy due to artifacts in calculation approximations.

|  | Skin | | body | |
|---|---|---|---|---|
|  | MD | DFT | MD | DFT |
| $d_H$ | 0.9576 | 0.9824 | 1.0002 | 1.0001 |
| $d_L$ | 1.7010 | 1.7966 | 1.9195 | 1.7732 |
| $q_O$ | - | -0.652 | - | -0.616 |
| $q_H$ | - | 0.314 | - | 0.319 |
| Net | - | -0.024 | - | 0.022 |

*4.4 Stress, viscosity, and elasticity*

Water and ice skins share high tension and viscosity. The surface tension γ is defined as the difference between the stress components in the direction parallel and perpendicular to the interface[62,63],

$$\gamma = \frac{1}{2}\left(\frac{\sigma_{xx}+\sigma_{yy}}{2} - \sigma_{zz}\right) \cdot L_z$$

(3)

where $\sigma_{xx}$, $\sigma_{yy}$, and $\sigma_{zz}$ are the stress tensor element and $L_z$ is the length of the super-cell in the *z* direction. Green [64] and Kubo [65] correlated the surface shear viscosity $\eta_s$ to the bulk stress σ in the way:

$$\eta_s = \frac{V}{kT}\int_0^\infty \langle \sigma_{\alpha\beta}(0)\sigma_{\alpha\beta}(t)\rangle dt$$

$$\text{(4)}$$

where the $\sigma_{\alpha\beta}$ denote the three equivalent off-diagonal elements of the stress tensors. The bulk viscosity $\eta_v$ is related to the decay of fluctuations in the diagonal elements of the stress tensor as follows:

$$\eta_v = \frac{V}{kT}\int_0^\infty \langle \delta\sigma(0)\delta\sigma(t)\rangle dt$$
$$\delta\sigma = \sigma - \langle\sigma\rangle$$

$$\text{(5)}$$

Based on these notations, we calculated the $\gamma$ using the MD-derived stress tensors. The auto-correlation functions of stress tensors can also be calculated and hence $\eta_s$ and $\eta_v$ are obtained according to eq (4) and (5). Table 3 features the MD-derived thickness-dependent $\gamma$, $\eta_s$, and $\eta_v$ of ice skin. Reduction of the number of water molecule layers increases the values of $\gamma$, $\eta_s$, and $\eta_v$, which agrees with our expectations. The O:H-O cooperative relaxation and the associated entrapment and polarization enhance the stress tensors to reach the values of 73.6 for 5 layers, approaching the measured 72 dyn/cm at 25 °C.

Table 3 Thickness-dependent surface tension and viscosity of ice skin.

| Number-of-layer | 15 | 8 | 5 |
|---|---|---|---|
| $\gamma$ dyn/cm | 31.5 | 55.2 | 73.6 |
| $\eta_s$ | 0.0007 | 0.0012 | 0.0019 |
| $\eta_v$ | 0.0027 | 0.0029 | 0.0032 |

Skin supersolidity slipperizes ice. The H-O contraction, core electron entrapment and nonbonding electron polarization yield the high-elasticity, self-lubrication, and low-friction of ice and the hydrophobicity of water surface as well. The mechanism of frictionless and self-lubricate of ice [5] is the same to that of metal nitride surfaces [66]. The friction coefficient of ice and some nitrides are in the same order of 0.1 or below. The elastic recovery coefficient of TiCrN and GaAlN surfaces reaches 100% under a critical indentation load of friction (< 1.0 mN) at which the lone pair breaks. The involvement of lone pairs makes the ice and the nitride skins more elastic and slippery at loads under the critical values.

# 5 Summary

In summary, experimental and numerical results evidence consistently our skin supersolidity prediction and the proposed physical origin - O:H-O bond cooperative relaxation and the associated charge entrapment and polarization. Molecular undercoordination shortens and stiffens the H-O bond and meanwhile lengthens and polarizes the O:H bond. In the skin region, molecular becomes smaller but the molecular distance becomes longer and hence the density of the skin drops substantially (0.75 g·cm$^{-3}$). The shortening of the H-O bond raises the density of the core and the bond electrons, which in turn polarize the lone pairs, resulting in the high elasticity and the high density of dipoles. The supersolid skin slipperizes ice and enhances the surface tension of liquid. Neither liquid skin forms on ice nor ice skin forms on water.

**Acknowledgement**

Financial support received from NSF (Nos.: 21273191, 1033003, and 90922025) China and RG29/12, MOE, Singapore is gratefully acknowledged.

**Multimedia movie Available:** A high-speed video clip showing the skin supersolidity (elasticity and hydrophobicity) is available free of charge via the Internet at http://pubs.acs.org.

**Corresponding Author**

zhouyc@xtu.edu.cn; ecqsun@ntu.edu.sg; CQ is associated with honorary appointments at the rest affiliations.